\documentclass[conference]{IEEEtran}
\IEEEoverridecommandlockouts
\usepackage{cite}
\usepackage{amsmath,amssymb,amsfonts}
\usepackage{algorithmic}
\usepackage{graphicx}
\usepackage{textcomp}
\usepackage{xcolor}
\def\BibTeX{{\rm B\kern-.05em{\sc i\kern-.025em b}\kern-.08em
    T\kern-.1667em\lower.7ex\hbox{E}\kern-.125emX}}

\usepackage[acronyms,nonumberlist,nopostdot,nomain,nogroupskip]{glossaries}
\newacronym{quic}{QUIC}{Quick UDP Internet Connections}
\newacronym{3gpp}{3GPP}{3rd Generation Partnership Project}
\newacronym{adc}{ADC}{Analog to Digital Converter}
\newacronym{5g}{5G}{5th Generation}
\newacronym{5gaa}{5GAA}{5G Automotive Association}
\newacronym{aimd}{AIMD}{Additive Increase Multiplicative Decrease}
\newacronym{am}{AM}{Acknowledged Mode}
\newacronym{amc}{AMC}{Adaptive Modulation and Coding}
\newacronym{aqm}{AQM}{Active Queue Management}
\newacronym{awgn}{AGWN}{Additive White Gaussian Noise}
\newacronym{balia}{BALIA}{Balanced Link Adaptation}
\newacronym{bdp}{BDP}{Bandwidth-Delay Product}
\newacronym{bf}{BF}{Beamforming}
\newacronym{cc}{CC}{Congestion Control}
\newacronym{pdf}{PDF}{Probability Density Function}
\newacronym{cdf}{CDF}{Cumulative Distribution Function}
\newacronym{icdf}{Inverse CDF}{Inverse Cumulative Distribution Function}
\newacronym{c-v2x}{C-V2X}{Cellular Vehicle-To-Everything}
\newacronym{ci}{CI}{Close-in free space reference}
\newacronym{cn}{CN}{Core Network}
\newacronym{cqi}{CQI}{Channel Quality Information}
\newacronym{cp}{CP}{Control Plane}
\newacronym{csirs}{CSI-RS}{Channel State Information - Reference Signal}
\newacronym{d2d}{D2D}{Device-to-Device}
\newacronym{dc}{DC}{Dual Connectivity}
\newacronym{dce}{DCE}{Direct Code Execution}
\newacronym{dci}{DCI}{Downlink Control Information}
\newacronym{dl}{DL}{downlink}
\newacronym{dmr}{DMR}{Deadline Miss Ratio}
\newacronym{dmrs}{DMRS}{DeModulation Reference Signal}
\newacronym{dsrc}{DSRC}{Dedicated Short-Range Communication}
\newacronym{e2e}{E2E}{end-to-end}
\newacronym{ecn}{ECN}{Explicit Congestion Notification}
\newacronym{edf}{EDF}{Earliest Deadline First}
\newacronym{enb}{eNB}{evolved Node Base}
\newacronym{epc}{EPC}{Evolved Packet Core}
\newacronym{es}{ES}{Edge Server}
\newacronym{fdma}{FDMA}{Frequency Division Multiple Access}
\newacronym{fdd}{FDD}{Frequency Division Duplexing}
\newacronym[firstplural=Radio Access Technologies (RATs)]{rat}{RAT}{Radio Access Technology}
\newacronym{fs}{FS}{Fast Switching}
\newacronym{ftp}{FTP}{File Transfer Protocol}
\newacronym{gnb}{gNB}{Next Generation Node Base}
\newacronym{harq}{HARQ}{Hybrid Automatic Repeat reQuest}
\newacronym{hetnet}{HetNet}{Heterogeneous Network}
\newacronym{hh}{HH}{Hard Handover}
\newacronym{hol}{HOL}{Head-of-Line}
\newacronym{ia}{IA}{Initial Access}
\newacronym{imt}{IMT}{International Mobile Telecommunication}
\newacronym{iot}{IoT}{Internet of Things}
\newacronym{lidar}{LiDAR}{Light Detection and Ranging}
\newacronym{los}{LOS}{Line of Sight}
\newacronym{lte}{LTE}{Long Term Evolution}
\newacronym{m2m}{M2M}{Machine to Machine}
\newacronym{mac}{MAC}{Medium Access Control}
\newacronym{mc}{MC}{Multi-Connectivity}
\newacronym{mcs}{MCS}{Modulation and Coding Scheme}
\newacronym{mec}{MEC}{Mobile Edge Cloud}
\newacronym{mi}{MI}{Mutual Information}
\newacronym{mimo}{MIMO}{Multiple Input, Multiple Output}
\newacronym{mmwave}{mmWave}{millimeter wave}
\newacronym{ml}{ML}{machine learning}
\newacronym{mr}{MR}{Maximum Rate}
\newacronym{mss}{MSS}{Maximum Segment Size}
\newacronym{mtd}{MTD}{Machine-Type Device}
\newacronym{mtu}{MTU}{Maximum Transmission Unit}
\newacronym{nn}{NN}{Neural Network}
\newacronym{nsf}{NSF}{National Science Foundation}
\newacronym{nfv}{NFV}{Network Function Virtualization}
\newacronym{nlos}{NLOS}{Non Line of Sight}
\newacronym{nr}{NR}{New Radio}
\newacronym{ofdm}{OFDM}{Orthogonal Frequency Division Multiplexing}
\newacronym{pc}{PC}{Point Cloud}
\newacronym{pdcch}{PDCCH}{Physical Downlonk Control Channel}
\newacronym{pdcp}{PDCP}{Packet Data Convergence Protocol}
\newacronym{pdsch}{PDSCH}{Physical Downlink Shared Channel}
\newacronym{pdu}{PDU}{Packet Data Unit}
\newacronym{pf}{PF}{Proportional Fair}
\newacronym{pgw}{PGW}{Packet Gateway}
\newacronym{phy}{PHY}{Physical}
\newacronym{pbch}{PBCH}{Physical Broadcast Channel}
\newacronym[plural=\gls{mme}s,firstplural=Mobility Management Entities (MMEs)]{mme}{MME}{Mobility Management Entity}
\newacronym{prb}{PRB}{Physical Resource Block}
\newacronym{pss}{PSS}{Primary Synchronization Signal}
\newacronym{pucch}{PUCCH}{Physical Uplink Control Channel}
\newacronym{pusch}{PUSCH}{Physical Uplink Shared Channel}
\newacronym{qos}{QoS}{Quality of Service}
\newacronym{rach}{RACH}{Random Access Channel}
\newacronym{ran}{RAN}{Radio Access Network}
\newacronym{red}{RED}{Random Early Detection}
\newacronym{rf}{RF}{Radio Frequency}
\newacronym{rlc}{RLC}{Radio Link Control}
\newacronym{rlf}{RLF}{Radio Link Failure}
\newacronym{rrc}{RRC}{Radio Resource Control}
\newacronym{rrm}{RRM}{Radio Resource Management}
\newacronym{rr}{RR}{Round Robin}
\newacronym{rs}{RS}{Remote Server}
\newacronym{rsrp}{RSRP}{Reference Signal Received Power}
\newacronym{rss}{RSS}{Received Signal Strength}
\newacronym{rtt}{RTT}{Round Trip Time}
\newacronym{rw}{RW}{Receive Window}
\newacronym{rx}{RX}{Receiver}
\newacronym{sa}{SA}{standalone}
\newacronym{sack}{SACK}{Selective Acknowledgment}
\newacronym{sap}{SAP}{Service Access Point}
\newacronym{sch}{SCH}{Secondary Cell Handover}
\newacronym{scoot}{SCOOT}{Split Cycle Offset Optimization Technique}
\newacronym{sdma}{SDMA}{Spatial Division Multiple Access}
\newacronym{sinr}{SINR}{Signal to Interference plus Noise Ratio}
\newacronym{sm}{SM}{Saturation Mode}
\newacronym{snr}{SNR}{Signal to Noise Ratio}
\newacronym{son}{SON}{Self-Organizing Network}
\newacronym{ss}{SS}{Synchronization Signal}
\newacronym{srs}{SRS}{Sounding Reference Signal}
\newacronym{sss}{SSS}{Secondary Synchronization Signal}
\newacronym{tb}{TB}{Transport Block}
\newacronym{tcp}{TCP}{Transmission Control Protocol}
\newacronym{udp}{UDP}{User Datagram Protocol}
\newacronym{fr1}{FR1}{Frequency Range 1}
\newacronym{fr2}{FR2}{Frequency Range 2}
\newacronym{tdd}{TDD}{Time Division Duplexing}
\newacronym{tdma}{TDMA}{Time Division Multiple Access}
\newacronym{tfl}{TfL}{Transport for London}
\newacronym{thz}{THz}{Terahertz}
\newacronym{tm}{TM}{Transparent Mode}
\newacronym{trp}{TRP}{Transmitter Receiver Pair}
\newacronym{tti}{TTI}{Transmission Time Interval}
\newacronym{ttt}{TTT}{Time-to-Trigger}
\newacronym{tx}{TX}{Transmitter}
\newacronym{ue}{UE}{User Equipment}
\newacronym{ul}{UL}{uplink}
\newacronym{uml}{UML}{Unified Modeling Language}
\newacronym{um}{UM}{Unacknowledged Mode}
\newacronym{utc}{UTC}{Urban Traffic Control}
\newacronym{v2v}{V2V}{Vehicle-to-Vehicle}
\newacronym{vm}{VM}{Virtual Machine}
\newacronym{rsrq}{RSRQ}{Reference Signal Received Quality}
\newacronym{rssi}{RSSI}{Received Signal Strength Indicator}
\newacronym{crs}{CRS}{Cell Reference Signal}
\newacronym{comp}{CoMP}{Coordinated Multi-Point}
\newacronym{cran}{C-RAN}{Cloud \acrlong{ran}}
\newacronym{ca}{CA}{Carrier Aggregation}
\newacronym{cco}{CC}{Carrier Component}
\newacronym{nsa}{NSA}{Non Stand Alone}
\newacronym{embb}{eMBB}{Enhanced Mobility Broadband}
\newacronym{bsr}{BSR}{Buffer Status Report}
\newacronym{srb}{SRB}{Service Radio Bearer}
\newacronym{scm}{SCM}{Spatial Channel Model}
\newacronym{sctp}{SCTP}{Stream Control Transmission Protocol}
\newacronym{mptcp}{MPTCP}{Multi-path TCP}
\newacronym{ietf}{IETF}{Internet Engineering Task Force}
\newacronym{os}{OS}{Operating System}
\newacronym{tls}{TLS}{Transport Layer Security}
\newacronym{rfc}{RFC}{Request for Comments}
\newacronym{http}{HTTP}{HyperText Transfer Protocol}
\newacronym{nat}{NAT}{Network Address Translation}
\newacronym{api}{API}{Application Programming Interface}
\newacronym{rto}{RTO}{Retransmission Timeout}
\newacronym{psc}{PSC}{Public Safety Communication}
\newacronym{rpgm}{RPGM}{Reference Point Group Mobility}
\newacronym{ic}{IC}{Incident Command}
\newacronym{rsu}{RSU}{Road Side Unit}
\newacronym{uav}{UAV}{Unmanned Aerial Vehicle}
\newacronym{usa}{U.S.}{United States}
\newacronym{vr}{VR}{Virtual Reality}
\newacronym{iab}{IAB}{Integrated Access and Backhaul}
\newacronym{wlan}{WLAN}{Wireless Local Area Network}
\newacronym{cots}{COTS}{Commercial Off-the-Shelf}
\newacronym{fpga}{FPGA}{Field Programmable Gate Array}
\newacronym{rcn}{RCN}{Research Coordination Network}
\newacronym{abg}{ABG}{Alpha-Beta-Gamma}
\newacronym{fi}{FI}{Floating Intercept}
\newacronym{uas}{UAS}{Unmanned Aerial System}
\newacronym{gps}{GPS}{Global Positioning System}
\newacronym{a2g}{A2G}{air-to-ground}
\newacronym{a2a}{A2A}{air-to-air}
\newacronym{uma}{UMa}{Urban Macro}
\newacronym{umi}{UMi}{Urban Micro}
\newacronym{upa}{UPA}{Uniform Planar Array}
\newacronym{rma}{RMa}{Rural Macro}
\newacronym{inoo}{InOo}{Indoor Open Office}
\newacronym{ple}{PLE}{path loss exponent}
\newacronym{aoa}{AoA}{Angle of Arrival}
\newacronym{aod}{AoD}{Angle of Departure}
\newacronym{toa}{ToA}{Time of Arrival}
\newacronym{mpc}{MPC}{Multi-path Component}
\newacronym{cir}{CIR}{Channel Impulse Response}
\newacronym{rt}{RT}{Ray-tracing}
\newacronym{tc}{TC}{Time Cluster}
\newacronym{sl}{SL}{Spatial Lobe}
\newacronym{svd}{SVD}{Singular Value Decomposition}
\newacronym{6g}{6G}{sixth generation}
\newacronym{ns3}{ns-3}{Network Simulator 3}
\newacronym{fsc}{FS}{Fully Stochastic}
\newacronym{hbc}{HB}{Hybrid}
\newacronym{hpbw}{HPBW}{Half Power Beamwidth}
\newacronym{hsc}{HSC}{Hybrid Semantic Compression}
\newacronym{prr}{PRR}{Packet Receipt Rate}
\newacronym{v2x}{V2X}{Vehicle-To-Everything}

\newacronym{ai}{AI}{artificial intelligence}

\newacronym{pqos}{PQoS}{Predictive Quality of Service}
\newacronym{rl}{RL}{Reinforcement Learning}
\newacronym{marl}{MARL}{Multi-Agent Reinforcement Learning}
\newacronym{mdp}{MDP}{Markov Decision Process}
\newacronym{pomdp}{POMDP}{Partially Observable MDP}
\newacronym{dec-pomdp}{Dec-POMDP}{Decentralized POMDP}
\newacronym{fl}{FL}{federated learning}
\newacronym{sgd}{SGD}{Stochastic Gradient Descent}
\newacronym{ddqn}{DDQN}{Double Deep Q-Network}
\newacronym{dql}{DQL}{Double Q-Learning}
\newacronym{td}{TD}{teleoperated driving}
\newacronym{map}{mAP}{mean Average Precision}
\newacronym{kpi}{KPI}{Key Performance Indicator}
\newacronym{mab}{MAB}{Multi-Armed Bandit}
\newacronym{dsarsa}{DSARSA}{Deep SARSA}
\newacronym{sarsa}{SARSA}{State–Action–Reward–State–Action}
\newacronym{qoe}{QoE}{Quality of Experience}
\newacronym{ppo}{PPO}{Proximal Policy Optimization}
\newacronym{trpo}{TRPO}{Trust Region Policy Optimization}
\newacronym{gae}{GAE}{Generalized Advantage Estimation}
\newacronym{ctde}{CTDE}{centralized training with decentralized execution}
\newacronym{ippo}{IPPO}{Independent PPO}
\newacronym{mappo}{MAPPO}{Multi-Agent PPO}
\newacronym{pa}{PA}{Proportional Allocation}
\newacronym{ga}{GA}{Greedy Allocation}
\newacronym{drl}{DRL}{Deep Reinforcement Learning}
\newacronym{urllc}{URLCC}{Ultra-Reliable Low Latency Communication}

\usepackage{hyperref}

\usepackage[font=small]{caption}
\usepackage[font=footnotesize]{subcaption}
\usepackage{booktabs} 
\usepackage[utf8]{inputenc}
\usepackage{pgfplots}
\DeclareUnicodeCharacter{2212}{−}
\usepgfplotslibrary{groupplots,dateplot}
\usetikzlibrary{patterns,shapes.arrows}
\pgfplotsset{compat=newest}

\usepackage{notation}
\usepackage{multirow}

\usepackage{tabularx}
\usepackage{subcaption}
\usepackage{flushend}

\usepackage[most]{tcolorbox}
\tcbuselibrary{theorems}

\newtcbtheorem{Definitions}{Summary}%
{colframe=black,colback=white,fonttitle=\bfseries}{}

\usetikzlibrary{positioning}

\begin{document}

\title{Multi-Agent Reinforcement Learning Scheduling to Support Low Latency in Teleoperated Driving}

\author{\IEEEauthorblockN{Giacomo Avanzi, Marco Giordani, Michele Zorzi\medskip}
\IEEEauthorblockA{
Department of Information Engineering, University of Padova, Italy. \\
Email: \texttt{\{giacomo.avanzi,marco.giordani,michele.zorzi\}@dei.unipd.it}
\vspace{-0.5cm}\\
}

\thanks{This work was supported by the European Union under the Italian National Recovery and Resilience Plan (NRRP) Mission 4, Component 2, Investment 1.3, CUP C93C22005250001, partnership on “Telecommunications of the Future” (PE00000001 - program “RESTART”)}}

\maketitle
\begin{abstract}
The \gls{td} scenario comes with stringent \gls{qos} communication constraints, especially in terms of \gls{e2e} latency and reliability. In this context, \gls{pqos}, possibly combined with \gls{rl} techniques, is a powerful tool to estimate \gls{qos} degradation and react accordingly. 
For example, an intelligent agent can be trained to select the optimal compression configuration for automotive data, and reduce the file size whenever QoS conditions deteriorate. However, compression may inevitably compromise data quality, with negative implications for the \gls{td} application.
An alternative strategy involves operating at the \gls{ran} level to optimize radio parameters based on current network conditions, while preserving data quality.
In this paper, we propose \gls{marl} scheduling algorithms, based on \gls{ppo}, to dynamically and intelligently allocate radio resources to minimize \gls{e2e} latency in a TD scenario.
We evaluate two training paradigms, i.e., decentralized learning with local observations (IPPO) vs. centralized aggregation (MAPPO), in conjunction with two resource allocation strategies, i.e., proportional allocation (PA) and greedy allocation (GA).
We prove via ns-3 simulations that MAPPO, combined with GA, achieves the best results in terms of latency, especially as the number of vehicles increases. 

\end{abstract}

\glsresetall


\glsresetall

\begin{tikzpicture}[remember picture,overlay]
\node[anchor=north,yshift=-10pt] at (current page.north) {\parbox{\dimexpr\textwidth-\fboxsep-\fboxrule\relax}{
\centering\footnotesize This paper has been accepted for publication in the 2025 IEEE Vehicular Networking Conference (VNC). \textcopyright 2025 IEEE. \\
Please cite it as: G. Avanzi, M. Giordani, and M. Zorzi, ``Multi-Agent Reinforcement Learning Scheduling to Support Low Latency in Teleoperated Driving,'' in Proc. IEEE Vehicular Networking Conference (VNC), 2025.\\
}};
\end{tikzpicture}

\section{Introduction}

In \gls{6g} networks, massive amounts of data will be exchanged, with human communication accounting only for a minimal fraction of the traffic~\cite{giordani6g}. 
Notably, vehicular communication is expected to be a key protagonist of 6G, interconnecting vehicles with other vehicles, infrastructures, pedestrians, and networks. However, fully autonomous driving with no human interaction presents critical technical challenges~\cite{TD}. Therefore, the research community is focusing on \gls{td}, where a remote driver controls the vehicles based on measurements and observations generated by onboard sensors, such as high-resolution videocameras and \gls{lidar} sensors. 

The performance of the \gls{td} application strongly depends on the network conditions in which the vehicles are deployed. 
In particular, strict requirements must be satisfied in terms of \gls{qos}. 
According to \gls{5gaa} specifications, the service-level latency with the remote driver depends on the automation level and, for \gls{td}, should not exceed 50 ms in both \gls{ul} and \gls{dl}, while reliability ranges from 99\% to 99.999\%~\cite{5gaa}. 
However, transmitting large volumes of data may require bit rates of hundreds of megabits per second~\cite{lidar}, and ultimately create network congestion.
Moreover, unanticipated channel degradation may lead to critical safety risks and/or reliability issues for \gls{td} applications.

For this reason, \gls{pqos} was introduced as a mechanism to forecast and communicate potential \gls{qos} changes in the network, and undertake proper countermeasures to react accordingly~\cite{boban2021predictive}. 
Notably, PQoS can be based on \glspl{nn}, leveraging input features related to network conditions, resource availability, predicted mobility patterns, and/or other observations. 
Recently, \gls{rl} methods have been also investigated to implement \gls{pqos} in \gls{td} scenarios. For example, in our previous works~\cite{mason,bragato}, we proposed a PQoS framework to select the optimal compression level for \gls{lidar} data to minimize the \gls{e2e} latency. 
A \gls{ddqn} model was used as the predictor, even though in ~\cite{bragato2024federated} we explored several other \gls{rl} alternatives. 
However, compression might inevitably degrade the quality of the LiDAR data, 
and possibly compromise \gls{td} operations such as object detection and recognition.

PQoS can also dynamically optimize \gls{ran} parameters, such as the transmission power, the numerology, or the communication spectrum,
based on QoS estimates~\cite{boban2021predictive}. 
An advantage of this approach is that it focuses exclusively on network-level parameters, thereby preserving the integrity and quality of the transmitted data compared to other methods that rely on, for example, compression.
Notably, the \gls{ran} can be optimized at the scheduling level, e.g., based on the temporal evolution of the communication channel and the available resources.
In fact, existing 5G schedulers, such as \gls{rr}, proportional fair, earliest-deadline first, were not designed to handle time-sensitive traffic. In turn, \gls{drl}, along with its multi-agent extension, has emerged as a powerful tool to schedule resources in a time-varying and unpredictable environment like in vehicular networks~\cite{drl}. In \cite{drl1}, the authors proposed a new scheduler implementing a knowledge-based \gls{drl} algorithm to deal with time-sensitive traffic in 5G networks. A similar strategy, also based on \gls{drl}, was proposed in \cite{drl2} to support \gls{urllc}.


Along these lines, in this paper we propose, implement, and evaluate novel \gls{marl} algorithms to optimize \gls{qos} (specifically, minimize the \gls{e2e} latency in a TD scenario), without compromising the accuracy of data. Our approach operates at the RAN level by training local agents that optimize scheduling based on latency conditions and the available network capacity.
Specifically, we investigate two extensions of the \gls{ppo} algorithm: \gls{ippo}, where 
we optimize multiple decentralized independent agents using local observations, and \gls{mappo}, where a single centralized model is trained using data from all local agents.
Moreover, we compare a \gls{pa} approach, in which resources are distributed fairly based on some priority levels, and a \gls{ga} approach where the available resources are assigned  to the vehicle experiencing the most severe latency, at the expense of others.
The algorithms are evaluated via ns-3 simulations as a function of the number of vehicles and the size of the transmitted data.
The results demonstrate that \gls{mappo}, combined with \gls{ga}, gives the best results in terms of latency, and can maximize the number of vehicles that satisfy latency constraints.

The rest of the paper is organized as follows.
In Sec.~\ref{sec:model} we describe our system model.
In Sec.~\ref{sec:marl} we present our MARL resource allocation algorithms.
In Sec.~\ref{sec:performance} we illustrate our main simulation results.
In Sec.~\ref{sec:conclusions} we conclude the paper with suggestions for future work.

\section{System Model}
\label{sec:model}
In this section we describe our simulation scenario (Sec.~\ref{sub:scenario}) and optimization framework (Sec.~\ref{sub:pqos}).

\subsection{Simulation Scenario}
\label{sub:scenario}

Our simulation scenario, based on~\cite{mason}, consists of a \gls{gnb}, a remote host (i.e., the teleoperator or driving software), $N$ vehicles (i.e., the \glspl{ue}), and the following modules.

\paragraph{Network}
A wired channel interconnects the remote host with the \gls{gnb}. The \gls{gnb} communicates with the vehicles via the 5G \gls{nr} protocol stack, which is simulated based on the open-source \texttt{mmwave} module for ns-3~\cite{mmwave}.

\paragraph{Channel and mobility model} The mobility of the vehicles is simulated in Simulation of Urban MObility (SUMO)~\cite{sumo}, in an area of the city of Bologna. The wireless channel and the propagation loss model are computed via the GEMV2 simulator~\cite{gemv}, and channel traces are parsed in ns-3 to compute the received~power.

\paragraph{Application data}
Each vehicle is equipped with a \gls{udp} application transmitting \gls{lidar} point clouds at a frame rate $f$.\footnote{While future teleoperated cars will be equipped with several types of sensors, including radar, camera and \gls{lidar} sensors, in this work, without loss of generality, we focus on the transmission of \gls{lidar} perceptions.} Data is eventually compressed by Google Draco~\cite{draco}. Specifically, this software defines several compression configurations based on the number of quantization bits $q$ $\in \{1,..., 31\}$ and the number of compression levels $c$ $\in \{0,..., 10\}$.

\paragraph{RAN-AI}
\label{par:ran-ai}
The RAN-AI entity~\cite{ran-ai} is an intelligent network controller, installed in the \gls{gnb}.
Specifically, it collects measurements and metrics from the \gls{ran} (e.g., \gls{e2e} latency, \gls{sinr}, etc.), and optimizes network operations to satisfy QoS constraints. 
In our previous work, the RAN-AI was trained based on a single centralized~\cite{mason} or decentralized~\cite{bragato2024federated} \gls{rl} agent. 
Rather, in this paper we study and implement different multi-agent RL solutions, as described in Sec.~\ref{sec:marl}.

\subsection{Optimization Framework}
\label{sub:pqos}
\Gls{pqos} aims at anticipating communication impairments and taking proper countermeasures to avoid service degradation~\cite{boban2021predictive}. 
At the RAN level, these countermeasures include, for example,
adjusting data compression to reduce network congestion, adapting the periodicity and speed of data transmissions to ensure service reliability, and/or modifying vehicle speed and trajectory based on route predictions and conditions.
While our previous work focused on the application layer, in this paper we operate at the scheduler level, and optimize radio resource allocation to minimize the \gls{e2e} latency.

We exploit the flexibility of the frame structure in 5G networks.
According to the NR standard, the available time resources are arranged into frames of 10 ms, each of which consists of 10 subframes of 1 ms and a number of slots that depends on the selected numerology~\cite{38300}.
Each slot consists of 14 \gls{ofdm} symbols, assuming normal Cyclic Prefix (CP), 
whose duration also depends on the numerology.
In 5G NR with \gls{tdma}, dynamic downlink scheduling occurs at the OFDM symbol level, meaning that the scheduler can assign radio resources with the granularity of individual OFDM symbols within a time slot, rather than that of the full slot or the subframe as in 4G LTE. 
The RAN-AI entity, described in Sec.~\ref{sub:scenario}, implements an \gls{marl} algorithm that 
determines the optimal number of resources, i.e., OFDM symbols, to be allocated to each UE for transmission.

Channel resources are limited, and all UEs compete for those resources to satisfy latency constraints.
Notably, radio resource allocation is governed by a priority level parameter $k \in \{1, \dots, K \}$, which is related to the network conditions of a given UE. 
Hence, lower-priority UEs ($k\rightarrow 1$) receive fewer resources, as latency requirements are easier to satisfy, while higher-priority UEs ($k \rightarrow K$) require more resources.
The number of priority levels $K$ defines the granularity of the agent’s decision.
We consider two scheduling methodologies, namely a proportional and a greedy approach (as described in Sec. \ref{sec:marl}). 
A baseline \gls{rr} scheduler is used as our benchmark.

\section{Proposed \gls{marl} Scheduling Algorithm for \gls{td}}\label{sec:marl}

\gls{rl} is a \gls{ml} technique where an agent interacts with the environment to learn how to maximize a cumulative future reward. The \gls{rl} framework can be formalized mathematically as a \gls{mdp}, defined by the tuple $<\mathcal{S}, \mathcal{A}, \mathcal{P}, \mathcal{R}, \mathcal{\gamma}>$ such that $\mathcal{S}$ is the finite set of states, $\mathcal{A}$ is the finite set of actions, $\mathcal{P}$ is the state transition probability matrix with elements $\mathcal{P}^a_{ss'} = P[S_{t+1}=s' | S_t=s, A_t=a]$, $\mathcal{R}$ is the reward function with $\mathcal{R}^a_s = E[R_{t+1} | S_t=s, A_t=a]$, and $\gamma \in [0, 1)$ is the discount factor.
More precisely, at each time step $t$, the agent interacts with the environment, observes the state $S_t$, takes an action $A_t$, receives a reward $R_{t+1}$, and moves to state $S_{t+1}$ according to $\mathcal{P}$. 
The goal of the agent is to find the optimal policy $\pi^*$ that maximizes the infinite-horizon expected return $G_t$, defined as the sum of the discounted rewards from time $t$. Specifically, $G_t$ is defined as
\begin{equation}
	G_t=\sum_{\tau=0}^{+ \infty} \gamma^\tau R_{t+\tau}.
\end{equation}

In the case of a \gls{pomdp}, the agent only perceives an observation $O_t$ of $S_t$, 
which provides partial information about the underlying state $S_t$.

Various algorithms have been developed to determine $\pi^*$. 
While in our previous work we focused on single-agent policy-based RL algorithms, in this paper we extend the analysis to consider a multi-agent approach (MARL), as described in Sec.~\ref{sub:model}.
Then, in Secs.~\ref{sub:marl-algorithms} and~\ref{sub:scheduling} we present our MARL and scheduling algorithms, respectively.

\subsection{Formalization of the Model}
\label{sub:model}
A centralized \gls{marl} problem is characterized by $N$ agents, where each agent aims at maximizing its own total expected return, while interacting with the other agents in a dynamic environment . 
However, in this approach, the size of the action space grows exponentially with the number of agents~\cite{marl}. 
To address this challenge, the problem is decomposed into a smaller and more tractable decentralized decision problem.
Specifically, we consider a \gls{dec-pomdp}~\cite{dec-pomdp}, i.e., a multi-agent extension of a \gls{pomdp}. It is defined as a tuple $<\mathcal{N}, \mathcal{S}, \{\mathcal{A}_i \}_{i \in \mathcal{N}}, \{\mathcal{O}_i \}_{i \in \mathcal{N}}, \mathcal{P}, \mathcal{R}, \mathcal{\gamma}>$, where $\mathcal{N}$ is a finite set of agents, $\mathcal{S}$ is a finite set of states, $\mathcal{A}_i$ is the finite set of actions for agent $i \in \mathcal{N}$, $\mathcal{O}_i$ is the finite set of observations for agent $i \in \mathcal{N}$, $\mathcal{P} :\mathcal{S} \times \mathcal{A}_\mathcal{N} \times \mathcal{S} \rightarrow \left[0, 1\right]$ is a state transition probability function, and $\mathcal{R}: \mathcal{S} \times \mathcal{A}_\mathcal{N} \rightarrow \mathbb{R}$ is the reward function. 
Each agent learns its decentralized policy, utilizing only its local observations and  rewards, while interacting with the shared environment.

Notably, the RAN-AI entity described in Sec.~\ref{sub:scenario} can be modeled as an \gls{marl} problem, and framed into a \gls{dec-pomdp} model. We provide the following definitions of state $\mathcal{S}$, observation $\mathcal{O}$, action $\mathcal{A}$ and reward $\mathcal{R}$.

\paragraph{State and observation} 
The state (observation) is defined as a set of network measurements from all UEs (from a single UE). These measurements are gathered by the RAN-AI in the \gls{gnb} through dedicated control signals during data transmissions. Specifically, the state/observation consists of the following metrics: the average \gls{sinr}, the 
\gls{ul} buffer size, the number of OFDM symbols required to transmit the data in the \gls{ul} buffer (given the \gls{mcs}), the average \gls{mcs} index, and the average \gls{e2e} latency and number of bytes transmitted at the application~layer.

\paragraph{Action} The action space $\mathcal{A}_i$ is identical for every UE$_i$, $i \in \mathcal{N}$. The action is defined as a scalar value $k \in \{1, ..., K\}$ corresponding to the priority level assigned to UE$_i$ at each resource allocation opportunity (see Sec.~\ref{sub:pqos}).

\paragraph{Reward} The reward function is designed to indicate if latency requirements are satisfied by a certain UE. 
Specifically, a positive reward is returned if the \gls{e2e} latency  $\ell$ at the application layer is lower than or equal to a predefined threshold $\tau$; otherwise, the reward is a penalization proportional to the violation of $\tau$. So, the reward function is defined as:
\begin{equation}
    R = \begin{cases}
        1&\text{if } \ell \leq \tau,\\
        -{(\ell - \tau)}/{100}&\text{otherwise}. \\
    \end{cases}
\end{equation}

\subsection{MARL Algorithms}
\label{sub:marl-algorithms}
We consider a \gls{ppo} algorithm for the training of the \mbox{RAN-AI}~\cite{ppo} since, contrary to more traditional methods like Q-learning, 
it is more suitable to manage non-stationary multi-agent environments as the size of the network, i.e., the number of agents/UEs, increases.
Specifically, \gls{ppo} is a model-free method derived from the \gls{trpo} algorithm~\cite{trpo}, which alternates between interaction with the environment and optimization (in multiple epochs) of a clipped surrogate objective function using \gls{sgd}. 
Let $r_t(\theta) = {\pi_\theta (a_t | o_t)}/{\pi_{\theta_{o}} (a_t | o_t)}$ be the probability ratio measuring the divergence between an updated parameterized policy $\pi_\theta$ and the original policy  $\pi_{\theta_{{o}}}$ (i.e., before the most recent parameters update). Then, let $\hat{A}_t$ be an estimator of the advantage function at time $t$, defined as the difference between the state-action value function (i.e., the $Q$-function) and the state value function. 
The clipped objective function can be written as
\begin{equation}
L^{\text{CL}}(\theta) \!=\! \mathbb{E}_t\!\left[ \min \left( r_t(\theta) \hat{A}_t, \, \!\text{clip}\left(r_t(\theta),\!1-\epsilon,\!1+\epsilon\right)\!\hat{A}_t \right) \right]
\end{equation}
where $\epsilon$ is an hyperparameter.

Moreover, since in PPO a state value function approximator $V_\phi$ is implemented and exploration is encouraged, the final objective function becomes
\begin{equation}
L(\theta, \phi) = L^{\text{CL}}(\theta) - c_1 L^{\text{VF}}(\phi) + c_2 S(\pi_\theta),
\end{equation}
where $L^{\text{VF}}$ is the mean squared error between $V_\phi$ 
and the target return $G_t$, and $S(\pi_\theta)$ represents the entropy of the policy. Constants $c_1$ and $c_2$ are hyperparameters that balance the contribution of the two terms.

The implementation of this model involves two \glspl{nn}, i.e., a policy network $\pi_\theta$ (actor) and a value network $V_\phi$ (critic). The former represents the policy of the agent; indeed, it receives as input the state and gives as output a probability distribution over the action space. The latter represents the state value function, and contributes to reducing the variance of the advantage function, i.e., of the gradient estimates. 
The NNs are fully connected: for $\pi_\theta$, we have 
$|\mathcal{O}_i|$ input neurons and 
$|\mathcal{A}_i|$ output neurons; for $V_\phi$, we have $|\mathcal{O}_i|$ input neurons and a single output neuron. 
There are two fully-connected hidden layers with $n_N$ neurons each, using the hyperbolic tangent as activation function, except for the output layer of $\pi_\theta$ where the softmax function is adopted. The parameters of the \glspl{nn} are updated using the Adam algorithm, with a learning rate $\alpha$. During the training, PPO is executed to generate trajectories of a fixed length of $T$ steps, which are tuples of states, actions and rewards collected interacting with the environment. During the learning, these trajectories are split into mini-batches of size $M$ to compute the gradient for improving the stability.

The \gls{gae}~\cite{gae} technique is used to approximate the advantage function $\hat{A}_t$. 
Notably, we use parameter $\lambda \in [0, 1]$ to control the trade-off between bias (due to systematic errors in the estimation of $\hat{A}_t$) and variance (due to noise in long trajectories). 
Formally, the advantage function at time $t$ is computed as
\begin{equation}
    \hat{A}_t = \sum_{l=0}^{T-t} (\gamma \lambda)^l \delta_{t+l} \text{ ,}
\end{equation}
where $\delta_{t} = r_{t} + \gamma V(s_{t+1}) - V(s_{t})$ is the temporal difference error at time $t$. 

In this paper, we explore two \gls{ppo} implementations. 

\paragraph{\acrfull{ippo}}\gls{ippo}~\cite{ippo} is the multi-agent version of \gls{ppo} where $N$ decentralized and independent policies are learnt by using only local observations. 
Therefore, the \gls{marl} problem involving $N$ agents is decomposed into $N$ single-agent problems. 
This approach is very effective and scalable, but does not guarantee learning stability or convergence to the optimal policy. 
In fact, from the point of view of an agent, the simultaneous learning process of the other agents introduces additional dynamics that may compromise the stationarity of the environment. 
\paragraph{\acrfull{mappo}}\gls{mappo}~\cite{mappo} is an example of a \gls{ctde} framework in which model parameters are shared to efficiently collect information in a centralized fashion~\cite{ctde}. 
Instead of having isolated agents, this approach updates a single actor and a single critic using data gathered from all $N$ agents. Therefore, all agents share the same policy and value function network. This technique accelerates the learning, is easy to implement, and more scalable with the number of agents than other \gls{ctde} approaches~\cite{ctde}. However, since observations come from multiple agents, the estimates of the advantage function have a high variance, making the system unstable and more difficult to generalize.


\subsection{Scheduling Algorithms}
\label{sub:scheduling}
Our scheduling approach is to allocate, for each UE, a certain number of OFDM symbols per slot based on the priority level $k$ (i.e., the action of the \gls{marl} algorithm based on IPPO or MAPPO). Notably, we implement two strategies.

\paragraph{\acrfull{pa}} 
The number of OFDM symbols $u_i$ allocated to UE$_i$, $i \in \{1,\dots,N\}$, is computed as
\begin{equation}
    u_i =\left \lfloor{U\frac{k_i}{\sum_{i\in \mathcal{N}}{k_i}}}\right \rfloor,
\end{equation} 
where $U$ is the number of available OFDM symbols/slot.
If $u_i<U$, 
the remaining OFDM symbols in the slot are used to serve other UE transmissions, starting from the UE(s) with the highest priority. 
Therefore, a principle of fairness is~preserved.

\paragraph{\acrfull{ga}} The allocation of OFDM symbols within a slot is greedy with respect to the priority level. Specifically, all $U$ symbols are assigned to the UE with the absolute highest priority. 
Unallocated symbols, if any, are assigned to the next UE(s) with higher priority. This procedure is repeated iteratively until the slot is completely allocated.

For both PA and GA, the allocation of resources is upper bounded by the number of symbols required to transmit the actual content (data) of the buffer of each UE, given the \gls{mcs}.

\section{Performance Evaluation}
\label{sec:performance}

\begin{table}
    \centering
    \caption{Simulation parameters.}
    \label{tab:simulation_parameters}
    \begin{tabular}{l|l}
        \hline
        Parameter & Value\\
        \hline
       Number of UEs/agents ($N$) & $\{3,  5, 8 \}$\\
       Carrier frequency ($f_c$) & 28 GHz \\
       Bandwidth ($B$) & 50 MHz \\
       Available OFDM symbols/slot ($U$) & 12 \\
        LiDAR frame rate ($f$) & 30 fps\\
        Latency threshold ($\tau$) & $\{15, 25, 35\}$ ms\\
        Number of priority levels ($K$) & 3\\
        Discount factor ($\gamma$) & 0.95 \\
       GAE parameter ($\lambda$) & 0.95 \\
        Number of neurons in hidden layers ($n_N$) & 64 \\
        Learning rate ($\alpha$) & $10^{-4}$\\
        Hyperparameters ($\{\epsilon, c_1,  c_2\}$) & \{0.2, 0.5, 0.01\}\\
        Length of a trajectory ($T$) & 512 steps\\
        Mini-batch size ($M$) & 64 steps \\
        \hline
    \end{tabular}
\end{table}
In this section, we first describe our simulation parameters (Sec.~\ref{sub:params}), then we present our numerical results (Sec.~\ref{sub:results}).

\subsection{Simulation Parameters}
\label{sub:params}
Our simulation scenario is implemented in ns-3, a system-level, end-to-end,  scalable, and open-access simulator of wireless networks. Notably, ns-3 comes with a dedicated module to simulate and test \gls{ml}/RL algorithms within the \gls{ran}~\cite{ran-ai} based on the pipeline described in Sec.~\ref{sec:model}, that we extended to implement our MARL IPPO and MAPPO approaches.\footnote{Source code: \url{https://github.com/signetlabdei/ns3-ran-ai}.}
Simulation parameters are reported in Table~\ref{tab:simulation_parameters}, and described~below.

\paragraph{Communication}
We consider 5G NR communication between the gNB and the UE(s) at a carrier frequency $f_c$ of 28 GHz and with a bandwidth $B$ of 50 MHz, so as to maximize the channel capacity. The 5G NR slot consists of $U=12$ available OFDM symbols, given that the first 2 symbols are reserved for control in \gls{ul} and \gls{dl}. We use numerology 3, so the resulting OFDM symbol duration is $8.92$~$\mu$s.
The \gls{gnb} (UE) has a transmission power of 30 (23) dBm. The (ideal) wired channel has a propagation delay of 10 ms and a transfer data rate of 100 Gbps. 

\paragraph{Application}
We consider an application generating LiDAR point clouds at a rate $f=30$ fps. For simplicity, we restrict our analysis to a (representative) subset of Draco compression configurations $(q,c)$, with $q \in \{8, 9, 10\}$ and $c \in \{0, 5, 10\}$. 
Specifically, $(8, 0)$ is the most aggressive configuration, resulting in a compressed data size that is roughly half of the most conservative configuration $(10, 10)$.

\paragraph{Learning algorithm} 
For the policy network $\pi_\theta$, we consider 6 input neurons equal to the size of the state/observation space, and $K$ output neurons equal to the size of the action space, i.e., the number of scheduling priority levels.
We empirically set $K=3$ based on offline simulations: a smaller $K$ would be insufficient to properly differentiate UEs, while increasing $K$ could lead to a complex and/or unstable learning environment, especially when the number of priority levels approximates that of the UEs. 
Our \gls{marl} algorithms are trained on 250 episodes with 400 learning steps for every UE/agent. Each episode is an independent simulation in ns-3, where 400 transmissions of point clouds are performed. The rest of the learning parameters are reported in Table~\ref{tab:simulation_parameters}.

\paragraph{Benchmarks}
We compare the performance of IPPO vs. MAPPO, using either PA or GA for resource allocation, for a total of 4 combinations.
For comparison, we consider an \gls{rr} benchmark in which UEs are assigned the same number of resources, regardless of the priority level, so independent of the actual latency conditions.

\paragraph{Metrics}
We run 250 independent ns-3 simulations, and evaluate: (i) the average \gls{e2e} latency at the application layer, measured from the time at which a data packet is generated at the transmitter to the time it is received; (ii) the average reward over the episodes; and (iii) the average latency-success probability, that is the probability that the latency is lower than or equal to a threshold $\tau$, i.e., $P_{\ell \leq \tau}$.
We investigate the impact of the number of vehicles $N$, the compression configuration $(q, c)$, and the latency threshold $\tau$.

\subsection{Numerical Results}
\label{sub:results}

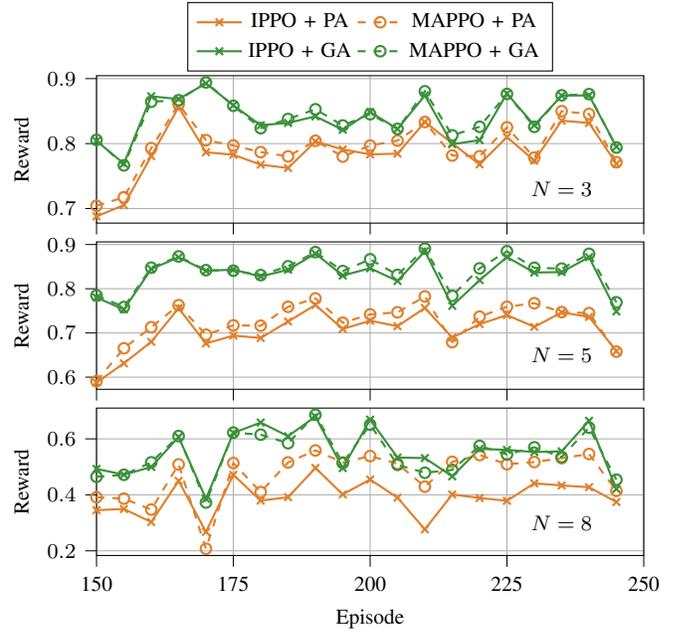
\begin{figure}
    \centering
\pgfplotsset{
tick label style={font=\footnotesize},
label style={font=\footnotesize},
legend  style={font=\footnotesize}
}
\begin{tikzpicture}

\definecolor{darkgray176}{RGB}{176,176,176}
\definecolor{darkorange25512714}{RGB}{255,127,14}
\definecolor{lightgray204}{RGB}{204,204,204}
\definecolor{steelblue31119180}{RGB}{31,119,180}
\definecolor{darkgray176}{RGB}{176,176,176}
\definecolor{darkslategray66}{RGB}{66,66,66}
\definecolor{darkslategray61}{RGB}{61,61,61}
\definecolor{lightgray204}{RGB}{204,204,204}
\definecolor{peru22412844}{RGB}{224,128,44}
\definecolor{seagreen5814558}{RGB}{58,145,58}
\definecolor{steelblue49115161}{RGB}{49,115,161}

\begin{groupplot}[
    group style={
        group size=1 by 3, 
        vertical sep=0.25cm, 
    },
    width=\columnwidth, 
    height=0.4\columnwidth 
]

\nextgroupplot[
legend style={at={(0.5,1.5)},
legend entries={IPPO + PA,
                MAPPO + PA,
                IPPO + GA,
                MAPPO + GA},
anchor=north,legend columns=2},
tick align=outside,
tick pos=left,
x grid style={darkgray176},
mark options={solid},
xmajorgrids,
xmin=30, xmax=50,
xtick={30, 35, 40, 45, 50},
xticklabels={150, 175, 200, 225, 250},
xtick style={color=black},
y grid style={darkgray176},
ymajorgrids,
ylabel={Reward},
ymin=0.677585206512532, ymax=0.904511311483213,
ytick style={color=black},
xticklabel=\empty 
]

\addlegendimage{mark=x, thick, peru22412844}
\addlegendimage{mark=o, dashed, thick, peru22412844}
\addlegendimage{mark=x, thick, seagreen5814558}
\addlegendimage{mark=o, dashed, thick, seagreen5814558}

\addplot [mark=x, thick, peru22412844, forget plot]
table {%
30 0.687900029465745
31 0.705301662076503
32 0.780657532357107
33 0.856298893628333
34 0.786657771408333
35 0.782839195051667
36 0.76755508455522
37 0.762429170331482
38 0.803226405161667
39 0.791055967388022
40 0.783442373311481
41 0.784565882917502
42 0.833242009311667
43 0.800564383734475
44 0.76790128151638
45 0.810136859918333
46 0.773691886165162
47 0.835291515185
48 0.831734966121667
49 0.768959457902168
};
\addplot [dashed, thick, mark=o, peru22412844, forget plot]
table {%
30 0.704203195357143
31 0.717344573626952
32 0.792965536573934
33 0.860704648975
34 0.805246375295
35 0.797022277456667
36 0.78690766562406
37 0.780313938279298
38 0.804359874781667
39 0.78004165352075
40 0.796953886488755
41 0.804545967957312
42 0.833284179466667
43 0.781627629598132
44 0.780514130050917
45 0.825065241685
46 0.77849523776648
47 0.849939366266667
48 0.845516567646667
49 0.771364890070458
};
\addplot [mark=x, thick, seagreen5814558, forget plot]
table {%
30 0.80414440831348
31 0.7690502683189
32 0.872764567828206
33 0.868582645421667
34 0.892659047258333
35 0.858914111608333
36 0.828190960073739
37 0.831853689768262
38 0.841853651036667
39 0.821350918397049
40 0.848706203385417
41 0.822581088696667
42 0.874009260706667
43 0.799014039645781
44 0.805375032797379
45 0.876910599868333
46 0.82821319207144
47 0.873863178056667
48 0.87423346366
49 0.79526691152439
};
\addplot [dashed, mark=o, thick, seagreen5814558, forget plot]
table {%
30 0.805736260575221
31 0.766501955905834
32 0.864439374564595
33 0.866771435661667
34 0.89419648853
35 0.858179921186667
36 0.823748715476429
37 0.837850591633065
38 0.852799381883333
39 0.827883453630862
40 0.845246475962251
41 0.822647768821651
42 0.880500134065
43 0.81275742065921
44 0.825878023485916
45 0.87645346493
46 0.8257898696356
47 0.874328329246667
48 0.876184198401667
49 0.793981342416027
};
\node [black] at (47, 0.73) {\footnotesize $N=3$};

\nextgroupplot[
legend style={at={(0.5,1.7)},
anchor=north,legend columns=-1},
tick align=outside,
tick pos=left,
x grid style={darkgray176},
mark options={solid},
xmajorgrids,
xmin=30, xmax=50,
xtick={30, 35, 40, 45, 50},
xticklabels={150, 175, 200, 225, 250},
xtick style={color=black},
y grid style={darkgray176},
ymajorgrids,
ylabel={Reward},
ymin=0.572556374555981, ymax=0.905947883970858,
ytick style={color=black},
xticklabel=\empty 
]
\vspace{-0.5cm}
\addplot [mark=x, thick, peru22412844, forget plot]
table {%
30 0.587710534074839
31 0.631355623087411
32 0.679976861988445
33 0.755885053133
34 0.676259900186485
35 0.694044274155256
36 0.688569956086687
37 0.725874553529351
38 0.763302951796
39 0.70895773084972
40 0.72766507879855
41 0.715312379122567
42 0.756503657108
43 0.689614865141289
44 0.719946264025586
45 0.74043664204
46 0.713766356719335
47 0.746457526164596
48 0.736267523079
49 0.658849400464669
};
\addplot [dashed, thick, mark=o, peru22412844, forget plot]
table {%
30 0.590996125272492
31 0.665239501349584
32 0.712954054749593
33 0.762791762886
34 0.695956883584117
35 0.717217871503751
36 0.716307336084243
37 0.759130549477131
38 0.778110473429
39 0.723110762943205
40 0.741967556771072
41 0.746684820471747
42 0.782591672641
43 0.679292686809026
44 0.736854960214551
45 0.759223064265
46 0.767230189166571
47 0.747141615707441
48 0.744826168877
49 0.657682774321159
};
\addplot [mark=x, thick, seagreen5814558, forget plot]
table {%
30 0.780227250627451
31 0.753084914825263
32 0.844591338473385
33 0.871149124215
34 0.840539758157625
35 0.843255923840459
36 0.827755363606442
37 0.842907183422268
38 0.879481117243
39 0.82992640141844
40 0.846364521220477
41 0.817489999195918
42 0.884826311581
43 0.761346101117175
44 0.819367948329771
45 0.871705772423
46 0.836601828716312
47 0.837689003296636
48 0.871364727121
49 0.748690337139159
};
\addplot [dashed, mark=o, thick, seagreen5814558, forget plot]
table {%
30 0.785013953804722
31 0.758662234175906
32 0.847610265274227
33 0.873071645917
34 0.842054838146646
35 0.8408656720052
36 0.830870667852301
37 0.850787547796731
38 0.882842977328
39 0.840006187604071
40 0.866577691974438
41 0.8312590859139
42 0.890793724452
43 0.784037635684894
44 0.846083764515601
45 0.884741489599
46 0.847475785303108
47 0.845455632230887
48 0.87898919243
49 0.769227432585959
};
\node [black] at (47, 0.65) {\footnotesize $N=5$};

\nextgroupplot[
legend style={at={(0.5,1.35)},
anchor=north,legend columns=-1},
tick align=outside,
tick pos=left,
x grid style={darkgray176},
mark options={solid},
xmajorgrids,
xlabel={Episode},
xmin=30, xmax=50,
xtick={30, 35, 40, 45, 50},
xticklabels={150, 175, 200, 225, 250},
xtick style={color=black},
y grid style={darkgray176},
ymajorgrids,
ylabel={Reward},
ymin=0.183379763927789, ymax=0.709936293043897,
ytick style={color=black}
]
\vspace{-0.5cm}
\addplot [mark=x, thick, peru22412844, forget plot]
table {%
30 0.345350269603752
31 0.349558329416139
32 0.302965633019082
33 0.449905278334884
34 0.267536749281143
35 0.471504793702781
36 0.379400606364183
37 0.391474255266462
38 0.496008102250542
39 0.401690804853774
40 0.45472529230984
41 0.389874739150552
42 0.276629571741012
43 0.401344158403396
44 0.388754218727934
45 0.378896069739264
46 0.441081674962359
47 0.433628485139535
48 0.427248409718811
49 0.374432747052557
};
\addplot [dashed, thick, mark=o, peru22412844, forget plot]
table {%
30 0.391360617953717
31 0.386590602295854
32 0.347417076525891
33 0.508793658081874
34 0.207314151614884
35 0.513517184683511
36 0.409803745558635
37 0.515778715363101
38 0.558627856317457
39 0.515633169310999
40 0.538507131599717
41 0.511179759578549
42 0.429331710792308
43 0.517917832892857
44 0.542161110041742
45 0.509409437431243
46 0.517271691620591
47 0.530885229765159
48 0.546025960970884
49 0.414194812570666
};
\addplot [mark=x, thick, seagreen5814558, forget plot]
table {%
30 0.492759243077273
31 0.471984706320034
32 0.499485380330595
33 0.606786637986686
34 0.385945067976779
35 0.621260026676865
36 0.658277736385468
37 0.60942641721864
38 0.679676961541309
39 0.495597033227691
40 0.668573724692865
41 0.533222023330153
42 0.531267726410643
43 0.466128953354552
44 0.564211407918118
45 0.56072635521149
46 0.552885869199045
47 0.555586937798137
48 0.664515420337617
49 0.420924933447412
};
\addplot [dashed, mark=o, thick, seagreen5814558, forget plot]
table {%
30 0.464531030707831
31 0.47160410934232
32 0.516126245613426
33 0.609903309084507
34 0.37264688363
35 0.622319478743961
36 0.614944636802575
37 0.58444632330458
38 0.686001905356801
39 0.513662082966255
40 0.650860875810694
41 0.507277345095509
42 0.478905368412729
43 0.488898451807044
44 0.575046910743844
45 0.544211473972486
46 0.569817323031777
47 0.535013276994382
48 0.64014263944887
49 0.454184950388307
};
\node [black] at (47, 0.3) {\footnotesize $N=8$};
\end{groupplot}

\end{tikzpicture}
    \vspace{-0.5cm}\caption{Average reward at the end of the training for \gls{ippo} and \gls{mappo}, combined with PA or GA, for $N \in \{3, 5,8\}$.\vspace{-0.cm}}
    \label{rewards_IPPO_MAPPO}
\end{figure}

\paragraph{Learning results} 
In Fig. \ref{rewards_IPPO_MAPPO}
we compare the learning performance of \gls{ippo} and \gls{mappo} in terms of reward.
As expected, \gls{mappo}, despite the increased complexity, generally achieves a higher reward than \gls{ippo} with both PA and GA scheduling options, given that the learning parameters are shared to a centralized node and optimized accordingly. Indeed, priorities for each vehicle are computed from a global perspective, resulting in a better coordination among the agents. 
More precisely, MAPPO demonstrates a more significant performance improvement in PA than in GA due to the inherently more complex nature of the former approach.
In GA, the learning is relatively straightforward, involving the identification of the most critical UE to allocate all of the available resources.
Conversely, PA requires the allocation of resources among multiple UEs based on their priority levels, which requires strong coordination. 
In this sense, the collaborative nature of MAPPO facilitates this coordination,
compared to an independent approach like IPPO, and can accelerate the convergence of the learning process for PA.

Moreover, the gap between IPPO and MAPPO increases as $N$ increases, especially for PA. Indeed, in crowded networks, the scheduling complexity increases, and requires more coordination among the agents to efficiently distribute resources, as promoted by MAPPO.

In view of the above results, in the rest of this section we continue our analysis considering only the \gls{mappo} algorithm.


\begin{figure}
    \centering
    \input{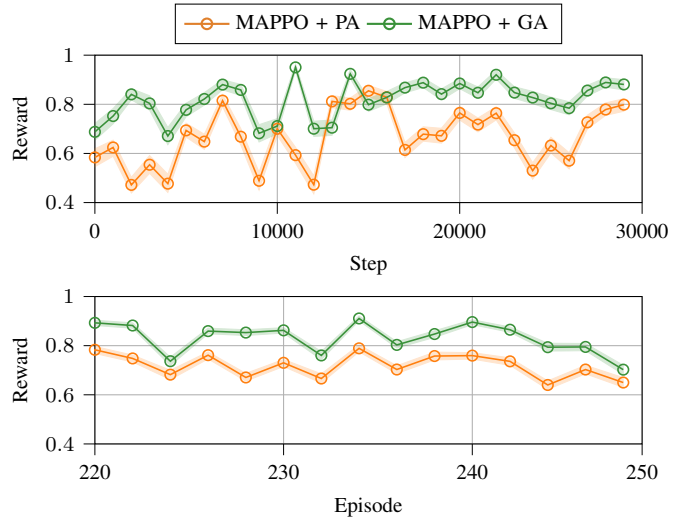}
    \vspace{-0.5cm}
    \caption{Average reward at the beginning (top) and at end of the training (bottom) for \gls{mappo}, with $N = 5$ .\vspace{-0.5cm}}
    \label{reward_MAPPO_5}
\end{figure}

In Fig. \ref{reward_MAPPO_5} we plot the evolution of the reward during the training of \gls{mappo}, focusing on the case of $N=5$.
In particular, at the beginning of the learning process, i.e., over the first 30k steps (15 episodes), the reward  improves for both PA and GA algorithms as the training progresses. Eventually, at the end of the training, i.e., in the last 30 episodes, MAPPO with GA achieves a higher reward (0.8) with less variance in comparison to the PA approach (0.7). 

\paragraph{Impact of the number of UEs}

In general, MAPPO outperforms a traditional \gls{rr} approach in terms of network performance. Specifically, we evaluate the average latency and the latency-success probability $P_{\ell\leq \tau}$.

    \begin{figure}
        \centering
        \input{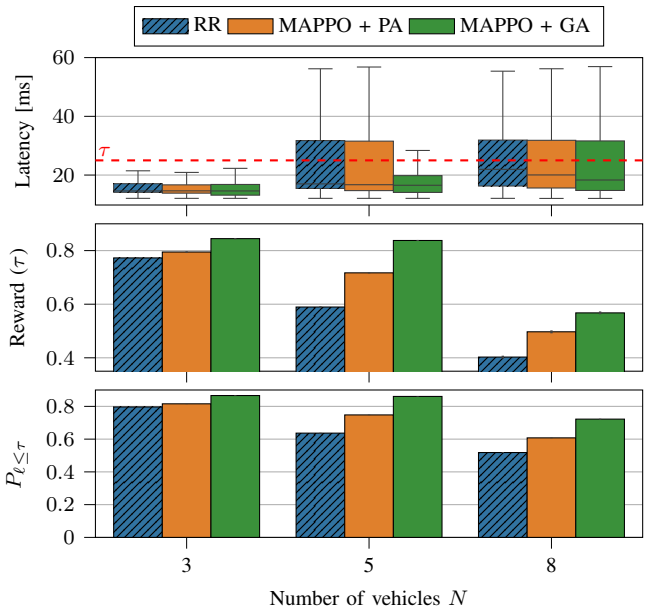}
        \caption{Average latency, reward and latency-success probability vs. $N$ at the end of the training, for $(q, c) = (8, 0)$ and $\tau$ = 25 ms.}
        \label{MAPPO_N}
    \end{figure}

In Fig.~\ref{MAPPO_N} we illustrate the average latency, reward, and the latency-success probability as a function of $N$ at the end of the training.
We observe that, while the median latency is always lower than $\tau$ with the current settings, its distribution depends on both $N$ and the scheduling approach. 
As $N$ increases, network congestion also increases, and so does the average (and variance of the) latency.
Specifically, with only $N=3$, the rewards for \gls{rr} and \gls{mappo} (with both \gls{pa} and \gls{ga} options) are very similar and close to 0.8.
 In this scenario, traffic requests can be easily handled, and network resources can be allocated without the need for coordination or more complex learning-based optimizations. In this sense, RR represents a simple and effective approach to support low~latency.
     
\begin{figure}
        \centering
        \pgfplotsset{compat=newest}
\usepgfplotslibrary{colorbrewer}
\pgfplotsset{
tick label style={font=\footnotesize},
label style={font=\footnotesize},
legend  style={font=\footnotesize}
}
\begin{tikzpicture}
    \begin{axis}[enlargelimits=false,
    colormap/Reds,
    tick align=outside,
    tick pos=left,
    xtick={0, 1, 2},
    xticklabels={3, 5, 8},
    xlabel = {Number of vehicles $N$},
    xtick style={color=black},
    ytick={0, 1, 2},
    yticklabel style={align=center},
    yticklabels={MAPPO\\+ GA, MAPPO\\+ PA, RR},
    y=1.1cm,
    ]
    \addplot [matrix plot,point meta=explicit]
      coordinates {
        (0,0) [0.42] (1,0) [0.36] (2,0) [0.45]

        (0,1) [0.25] (1,1) [0.31] (2,1) [0.27]

        (0,2) [0.32] (1,2) [0.33] (2,2) [0.27]
      };
    \node [black] at (0, 0) {\footnotesize 0.42};
    \node [black] at (1, 0) {\footnotesize 0.36};
    \node [white] at (2, 0) {\footnotesize 0.45};
    \node [black] at (0, 1) {\footnotesize 0.25};
    \node [black] at (1, 1) {\footnotesize 0.31};
    \node [black] at (2, 1) {\footnotesize 0.27};  
    \node [black] at (0, 2) {\footnotesize 0.32};
    \node [black] at (1, 2) {\footnotesize 0.33};
    \node [black] at (2, 2) {\footnotesize 0.27};

  \end{axis}
\end{tikzpicture}
        \caption{The 95-percentile of the latency-violation probability vs. $N$.\vspace{-0.3cm}}
        \label{p_violation}
    \end{figure}
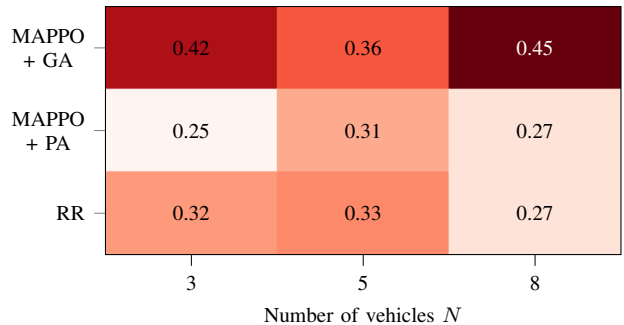

    Increasing $N$, and therefore the channel occupation, MAPPO consistently outperforms \gls{rr}, which demonstrates the benefits of MARL for resource allocation in a more complex, delay-critical scenario.
    For example, for $N=5$, the latency-success probability is around 35\% (15\%) higher with MAPPO using GA (PA) compared to the RR benchmark. This trend also appears from the boxplot in Fig.~\ref{MAPPO_N} (top), where RR and MAPPO with PA exhibit a significantly higher number of latency violations than MAPPO with GA, even though the median latency remains below $\tau$.
    In fact, as discussed, MAPPO with GA outperforms its PA counterpart as it prioritizes the most constrained UE (in terms of latency) by allocating more network resources, at the expense of the others.
    Meanwhile, \gls{pa} involves a principle of fairness by modulating the number of allocated resources to satisfy as many UE requests as possible. 
    Focusing on worst-case latency, in Fig.~\ref{p_violation} we plot the 95-th percentile of the latency-violation probability, i.e., $P_{\ell>\tau}$, vs. $N$, and as a function of the scheduling approach. We clearly see that MAPPO with GA yields the worst performance in this regard, since it aggressively prioritizes a limited subset of UEs,
    depriving others of sufficient channel resources. 
    Conversely, PA mitigates this negative condition, although it suffers from a higher average latency.
    
    Finally, as $N$ continues to increase, all scheduling solutions converge to similar average latency performance. In this congested scenario, the primary bottleneck of the system is represented by the limited number of resources, which are insufficient to accommodate all traffic requests, regardless of the underlying scheduling implementation.
    Nevertheless, MAPPO still outperforms RR in terms of reward and latency-success probability, and represents a more robust and scalable solution as $N$ increases. 

    \begin{figure}
        \centering
        \input{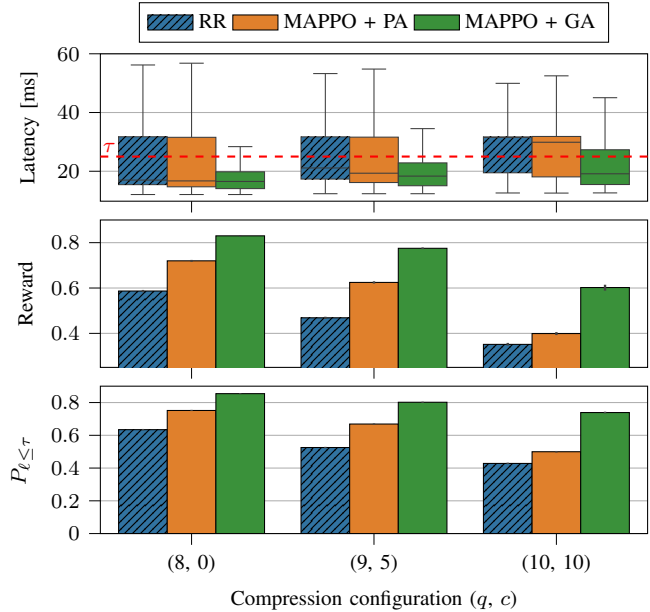}
        \caption{Average latency, reward and latency-success probability vs. the compression configuration at the end of the training, for $N$ = 5 and $\tau$ = 25 ms.\vspace{-0.5cm}}
        \label{MAPPO_q-c}
    \end{figure}

    \paragraph{Impact of the compression configuration}

     The compression configuration directly determines the size of the LiDAR point clouds to be transmitted, and therefore the data rate of the application. For example, for $(q, c) = (10, 10)$, the resulting aggregated data rate for $N=8$ is $66.2$ Mbps at a frame rate~$f = 30$ fps.
     At the end of the training, Fig. \ref{MAPPO_q-c} shows that the average reward is inversely proportional to the level of data compression,  dropping below 0.6 for $(q, c)=(10,10)$, even after optimization with MAPPO.
     This is because lower compression results in higher data rates, making latency constraints more difficult to satisfy.\footnote{Notice that, while higher compression reduces the data rate and network congestion, it may inevitably degrade the quality of data. This analysis is out of the scope of this paper, and was already partially addressed in~\cite{mason,bragato2024federated}.}
     Nevertheless, MAPPO consistently outperforms the benchmark \gls{rr} approach.
     Notably, under the most conservative compression configuration (i.e., $(q, c) =(10, 10)$), \gls{mappo} with \gls{ga} is the only approach capable of reducing the median latency below $\tau$, and satisfy latency constraints for more than 70\% of the time, vs. 50\% and 42\% for \gls{mappo} with \gls{pa} and RR, respectively.

    \begin{figure}
    \centering
\pgfplotsset{
tick label style={font=\footnotesize},
label style={font=\footnotesize},
legend  style={font=\footnotesize}
}
\begin{tikzpicture}

\definecolor{darkgray176}{RGB}{176,176,176}
\definecolor{darkslategray66}{RGB}{66,66,66}
\definecolor{lightgray204}{RGB}{204,204,204}
\definecolor{peru22412844}{RGB}{224,128,44}
\definecolor{seagreen5814558}{RGB}{58,145,58}
\definecolor{steelblue49115161}{RGB}{49,115,161}

\begin{axis}[
height=0.4\columnwidth,
width=\columnwidth,
legend style={at={(0.5,1.4)},
anchor=north,legend columns=-1},
tick align=outside,
tick pos=left,
x grid style={darkgray176},
xmin=-0.5, xmax=2.5,
xtick style={color=black},
xtick={0,1,2},
xticklabels={15,25,35},
xlabel={Latency threshold $\tau$ [ms]},
y grid style={darkgray176},
ylabel={Reward},
ymin=0, ymax=1.0,
ymajorgrids,
ytick style={color=black}
]
\draw[draw=black,fill=steelblue49115161, postaction={pattern=north east lines}] (axis cs:-0.4,0) rectangle (axis cs:-0.133333333333333,0.0701567113783316);
\draw[draw=black,fill=steelblue49115161, postaction={pattern=north east lines}] (axis cs:0.6,0) rectangle (axis cs:0.866666666666667,0.581186660441117);
\draw[draw=black,fill=steelblue49115161, postaction={pattern=north east lines}] (axis cs:1.6,0) rectangle (axis cs:1.86666666666667,0.90913270608737);
\draw[draw=black,fill=peru22412844] (axis cs:-0.133333333333333,0) rectangle (axis cs:0.133333333333333,0.0963357199607945);
\draw[draw=black,fill=peru22412844] (axis cs:0.866666666666667,0) rectangle (axis cs:1.13333333333333,0.700740355458809);
\draw[draw=black,fill=peru22412844] (axis cs:1.86666666666667,0) rectangle (axis cs:2.13333333333333,0.947467599439321);
\draw[draw=black,fill=seagreen5814558] (axis cs:0.133333333333333,0) rectangle (axis cs:0.4,0.290156957668211);
\draw[draw=black,fill=seagreen5814558] (axis cs:1.13333333333333,0) rectangle (axis cs:1.4,0.829876825534455);
\draw[draw=black,fill=seagreen5814558] (axis cs:2.13333333333333,0) rectangle (axis cs:2.4,0.941407773336179);

\draw[draw=none,fill=steelblue49115161] (axis cs:0,0) rectangle (axis cs:0,0);
\addlegendimage{area legend,draw=none,fill=steelblue49115161, postaction={pattern=north east lines}}
\addlegendentry{RR}

\draw[draw=none,fill=peru22412844] (axis cs:0,0) rectangle (axis cs:0,0);
\addlegendimage{area legend,draw=none,fill=peru22412844}
\addlegendentry{MAPPO + PA}

\draw[draw=none,fill=seagreen5814558] (axis cs:0,0) rectangle (axis cs:0,0);
\addlegendimage{area legend,draw=none,fill=seagreen5814558}
\addlegendentry{MAPPO + GA}

\addplot [line width=0.9pt, darkslategray66, forget plot]
table {%
-0.266666666666667 0.0689568982389873
-0.266666666666667 0.0715991176065038
};
\addplot [line width=0.9pt, darkslategray66, forget plot]
table {%
0.733333333333333 0.579446559501611
0.733333333333333 0.582731526886476
};
\addplot [line width=0.9pt, darkslategray66, forget plot]
table {%
1.73333333333333 0.908122660002356
1.73333333333333 0.910111016630148
};
\addplot [line width=0.9pt, darkslategray66, forget plot]
table {%
0 0.0951068637403317
0 0.0976532396133924
};
\addplot [line width=0.9pt, darkslategray66, forget plot]
table {%
1 0.699367276766408
1 0.702164412917959
};
\addplot [line width=0.9pt, darkslategray66, forget plot]
table {%
2 0.946655619850095
2 0.948243511801873
};
\addplot [line width=0.9pt, darkslategray66, forget plot]
table {%
0.266666666666667 0.288664413069321
0.266666666666667 0.291626240962116
};
\addplot [line width=0.9pt, darkslategray66, forget plot]
table {%
1.26666666666667 0.828716092928701
1.26666666666667 0.831107937723556
};
\addplot [line width=0.9pt, darkslategray66, forget plot]
table {%
2.26666666666667 0.940598873025014
2.26666666666667 0.94219175078846
};

\end{axis}

\end{tikzpicture}
    \caption{Average reward at the end of the training vs. $\tau$, for $N$ = 5 and $(q, c) = (8, 0)$.}
    \label{MAPPO_tau}
    \end{figure}
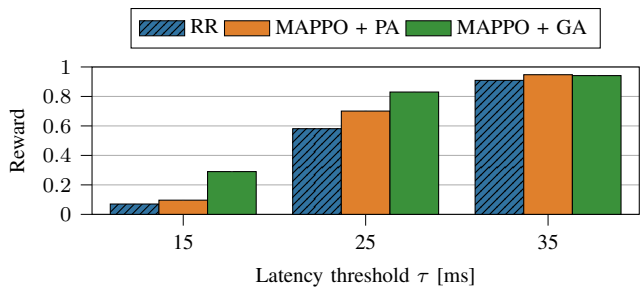

\paragraph{Impact of the latency threshold}
Finally, in Fig.~\ref{MAPPO_tau} we analyze the reward as a function of $\tau$.
In general, the reward decreases as $\tau$ decreases.
    For example, with $\tau = 15$ ms, \gls{mappo} with \gls{ga} outperforms RR by 0.2 ($+95\%$) in terms of reward, and stands out as the best scheduling approach in the most challenging environments. 
    With $\tau = 25$ ms, vehicles have more time to complete data transmission before violating the latency constraint, so the gap between MAPPO and RR is limited to $+35\%$. Still, \gls{mappo} with GA continues to be the best scheduler implementation, with an average reward around 0.8.
    As $\tau$ reaches $35$ ms, all scheduling options provide comparable performance 
    since the system is no longer heavily constrained, and resources can be efficiently allocated without requiring learning-based optimizations.
    This scenario is similar to the case of $N=3$ in Fig. \ref{reward_MAPPO_5}, where we demonstrated that, when the network is underutilized, MAPPO does not provide significant performance improvements over simpler approaches like RR.

\section{Conclusions and Future Work}
\label{sec:conclusions}

    In this work we proposed \gls{marl}-based scheduling algorithms to support low latency in a \gls{td} scenario in the context of \gls{pqos}.
    Specifically, we formalized a multi-agent model, and optimized the allocation of radio resources, i.e., OFDM symbols per slot, to minimize the probability of violating some predefined latency constraints.
    We evaluated two multi-agent extensions of \gls{ppo}, i.e., \gls{mappo} and \gls{ippo}, together with a proportional and a greedy strategy to distribute OFDM symbols based on priority levels.  
    To assess the performance of our proposed scheduler models, we run a simulation campaign in ns-3, using \gls{rr} as a benchmark. We showed that \gls{mappo} outperforms \gls{ippo} due to the fact that agents coordinate and share measurements with the network during the learning phase. 
     In particular, \gls{mappo} with \gls{ga} achieves lower average latency than its PA counterpart, and stands out as the most robust and effective scheduling approach, especially in the most constrained network configurations, e.g., when the number of vehicles and/or the application data rate increase. However, many severe latency violations are experienced, while \gls{pa} promotes fairness in resources allocation.

    As part of our future work, we plan to further extend our MARL framework by incorporating additional parameters, including the impact of compression on data quality and energy consumption, together with performance metrics such as throughput and latency.

\bibliography{ref}{}

\begin{thebibliography}{10}
\providecommand{\url}[1]{#1}
\csname url@samestyle\endcsname
\providecommand{\newblock}{\relax}
\providecommand{\bibinfo}[2]{#2}
\providecommand{\BIBentrySTDinterwordspacing}{\spaceskip=0pt\relax}
\providecommand{\BIBentryALTinterwordstretchfactor}{4}
\providecommand{\BIBentryALTinterwordspacing}{\spaceskip=\fontdimen2\font plus
\BIBentryALTinterwordstretchfactor\fontdimen3\font minus \fontdimen4\font\relax}
\providecommand{\BIBforeignlanguage}[2]{{%
\expandafter\ifx\csname l@#1\endcsname\relax
\typeout{** WARNING: IEEEtran.bst: No hyphenation pattern has been}%
\typeout{** loaded for the language `#1'. Using the pattern for}%
\typeout{** the default language instead.}%
\else
\language=\csname l@#1\endcsname
\fi
#2}}
\providecommand{\BIBdecl}{\relax}
\BIBdecl

\bibitem{giordani6g}
M.~Giordani, M.~Polese, M.~Mezzavilla, S.~Rangan, and M.~Zorzi, ``{Toward 6G Networks: Use Cases and Technologies},'' \emph{IEEE Communications Magazine}, vol.~58, no.~3, pp. 55--61, Mar. 2020.

\bibitem{TD}
T.~Zhang, ``Toward automated vehicle teleoperation: Vision, opportunities, and challenges,'' \emph{IEEE Internet of Things Journal}, vol.~7, no.~12, pp. 11\,347--11\,354, Dec. 2020.

\bibitem{5gaa}
5GAA, ``{C-V2X Use Cases Volume II: Examples and Service Level Requirements},'' White Paper, 2020.

\bibitem{lidar}
J.~Choi, V.~Va, N.~Gonzalez-Prelcic, R.~Daniels, C.~R. Bhat, and R.~W. Heath, ``Millimeter-wave vehicular communication to support massive automotive sensing,'' \emph{IEEE Communications Magazine}, vol.~54, no.~12, pp. 160--167, Dec. 2016.

\bibitem{boban2021predictive}
M.~Boban, M.~Giordani, and M.~Zorzi, ``{Predictive Quality of Service (PQoS): The Next Frontier for Fully Autonomous Systems},'' \emph{IEEE Network}, vol.~35, no.~6, pp. 104--110, Nov/Dec 2021.

\bibitem{mason}
F.~Mason, M.~Drago, T.~Zugno, M.~Giordani, M.~Boban, and M.~Zorzi, ``{A Reinforcement Learning Framework for PQoS in a Teleoperated Driving Scenario},'' in \emph{IEEE Wireless Communications and Networking Conference (WCNC)}, 2022.

\bibitem{bragato}
F.~Bragato, T.~Lotta, G.~Ventura, M.~Drago, F.~Mason, M.~Giordani, and M.~Zorzi, ``{Towards Decentralized Predictive Quality of Service in Next-Generation Vehicular Networks},'' in \emph{IEEE Information Theory and Applications Workshop (ITA)}, 2023.

\bibitem{bragato2024federated}
\BIBentryALTinterwordspacing
F.~Bragato, M.~Giordani, and M.~Zorzi, ``{Federated Reinforcement Learning to Optimize Teleoperated Driving Networks},'' in \emph{IEEE Global Communications Conference}, 2024. [Online]. Available: \url{https://arxiv.org/abs/2410.02312}
\BIBentrySTDinterwordspacing

\bibitem{drl}
L.~Liang, H.~Ye, G.~Yu, and G.~Y. Li, ``{Deep-Learning-Based Wireless Resource Allocation With Application to Vehicular Networks},'' \emph{Proceedings of the IEEE}, vol. 108, no.~2, pp. 341--356, Feb. 2020.

\bibitem{drl1}
Z.~Gu, C.~She, W.~Hardjawana, S.~Lumb, D.~McKechnie, T.~Essery, and B.~Vucetic, ``{Knowledge-Assisted Deep Reinforcement Learning in 5G Scheduler Design: From Theoretical Framework to Implementation},'' \emph{IEEE Journal on Selected Areas in Communications}, vol.~39, no.~7, pp. 2014--2028, Jul. 2021.

\bibitem{drl2}
Q.~Huang and M.~Kadoch, ``{5G Resource Scheduling for Low-latency Communication: A Reinforcement Learning Approach},'' in \emph{IEEE 92nd Vehicular Technology Conference (VTC2020-Fall)}, 2020.

\bibitem{mmwave}
M.~Mezzavilla, M.~Zhang, M.~Polese, R.~Ford, S.~Dutta, S.~Rangan, and M.~Zorzi, ``{End-to-End Simulation of 5G mmWave Networks},'' \emph{IEEE Communications Surveys \& Tutorials}, vol.~20, no.~3, pp. 2237--2263, Thirdquarter 2018.

\bibitem{sumo}
D.~Krajzewicz, J.~Erdmann, M.~Behrisch, and L.~Bieker, ``Recent development and applications of {SUMO - Simulation of Urban MObility},'' \emph{International Journal On Advances in Systems and Measurements}, vol.~5, no. 3\&4, pp. 128--138, Dec. 2012.

\bibitem{gemv}
M.~Boban, J.~Barros, and O.~K. Tonguz, ``Geometry-based vehicle-to-vehicle channel modeling for large-scale simulation,'' \emph{IEEE Transactions on Vehicular Technology}, vol.~63, no.~9, pp. 4146--4164, Nov. 2014.

\bibitem{draco}
\BIBentryALTinterwordspacing
Google. (2017) {Draco 3D Data Compression}. [Online]. Available: \url{https://google.github.io/draco/}
\BIBentrySTDinterwordspacing

\bibitem{ran-ai}
M.~Drago, T.~Zugno, F.~Mason, M.~Giordani, M.~Boban, and M.~Zorzi, ``{Artificial Intelligence in Vehicular Wireless Networks: A Case Study Using ns-3},'' in \emph{Proceedings of the 2022 ACM Workshop on Ns-3}, 2022.

\bibitem{38300}
{3GPP}, ``{NR and NG-RAN Overall Description (Release 15)},'' \emph{{TS 38.300}}, 2018.

\bibitem{marl}
S.~V. Albrecht, F.~Christianos, and L.~Sch\"afer, \emph{Multi-Agent Reinforcement Learning: Foundations and Modern Approaches}.\hskip 1em plus 0.5em minus 0.4em\relax MIT Press, 2024.

\bibitem{dec-pomdp}
D.~S. Bernstein, S.~Zilberstein, and N.~Immerman, ``The complexity of decentralized control of {M}arkov decision processes,'' in \emph{Proceedings of the Sixteenth Conference on Uncertainty in Artificial Intelligence}, 2000.

\bibitem{ppo}
\BIBentryALTinterwordspacing
J.~Schulman, F.~Wolski, P.~Dhariwal, A.~Radford, and O.~Klimov, ``Proximal policy optimization algorithms,'' 2017. [Online]. Available: \url{https://arxiv.org/abs/1707.06347}
\BIBentrySTDinterwordspacing

\bibitem{trpo}
J.~Schulman, S.~Levine, P.~Abbeel, M.~Jordan, and P.~Moritz, ``Trust region policy optimization,'' in \emph{Proceedings of the 32nd International Conference on Machine Learning}, 2015.

\bibitem{gae}
J.~Schulman, P.~Moritz, S.~Levine, M.~Jordan, and P.~Abbeel, ``High-dimensional continuous control using generalized advantage estimation,'' in \emph{Proceedings of the International Conference on Learning Representations (ICLR)}, 2016.

\bibitem{ippo}
\BIBentryALTinterwordspacing
C.~S. de~Witt, T.~Gupta, D.~Makoviichuk, V.~Makoviychuk, P.~H.~S. Torr, M.~Sun, and S.~Whiteson, ``Is independent learning all you need in the starcraft multi-agent challenge?'' 2020. [Online]. Available: \url{https://arxiv.org/abs/2011.09533}
\BIBentrySTDinterwordspacing

\bibitem{mappo}
C.~Yu, A.~Velu, E.~Vinitsky, J.~Gao, Y.~Wang, A.~Bayen, and Y.~Wu, ``{The surprising effectiveness of PPO in cooperative multi-agent games},'' in \emph{Proceedings of the 36th International Conference on Neural Information Processing Systems}, 2022.

\bibitem{ctde}
\BIBentryALTinterwordspacing
C.~Amato, ``An introduction to centralized training for decentralized execution in cooperative multi-agent reinforcement learning,'' 2024. [Online]. Available: \url{https://arxiv.org/abs/2409.03052}
\BIBentrySTDinterwordspacing

\end{thebibliography}
\bibliographystyle{IEEEtran}

\end{document}